\documentclass{aastex7}

\usepackage{amsmath}
\usepackage{amssymb}
\usepackage{graphicx}
\usepackage{alphabeta}

\begin{document}

\title{Disentangling the Effects of Temperature, Clouds, and Gravity on \ion{K}{1} doublet in L dwarfs}

\author[0009-0008-5619-4154]{Alexandra J. Baldelli}
\affiliation{Department of Astrophysics, American Museum of Natural History, Central Park West at 79th Street, NY 10024, USA}
\affiliation{Department of Applied Physics and Applied Mathematics, Columbia University, New York, NY, USA}
\email{Alexandra.J.Baldelli@gmail.com}

\author[0000-0002-2011-4924]{Genaro Su{\'a}rez}
\affiliation{Department of Astrophysics, American Museum of Natural History, Central Park West at 79th Street, NY 10024, USA}
\email{gsuarez@amnh.org}


\author[0000-0002-1821-0650]{Kelle L. Cruz}
\affiliation{Department of Astrophysics, American Museum of Natural History, Central Park West at 79th Street, NY 10024, USA}
\affiliation{Department of Physics and Astronomy, Hunter College, City University of New York, 695 Park Avenue, New York, NY 10065, USA}
\affiliation{Department of Physics, Graduate Center, City University of New York, 365 Fifth Avenue, New York, NY 10016, USA}
\email{kellecruz@gmail.com}

\author[0000-0001-6251-0573]{Jacqueline K. Faherty}
\affiliation{Department of Astrophysics, American Museum of Natural History, Central Park West at 79th Street, NY 10024, USA}
\affiliation{Department of Physics, Graduate Center, City University of New York, 365 Fifth Avenue, New York, NY 10016, USA}
\email{jfaherty@amnh.org}

\author[0000-0003-4083-9962]{Austin Rothermich}
\affiliation{Department of Astrophysics, American Museum of Natural History, Central Park West at 79th Street, NY 10024, USA}
\affiliation{Department of Physics and Astronomy, Hunter College, City University of New York, 695 Park Avenue, New York, NY 10065, USA}
\affiliation{Department of Physics, Graduate Center, City University of New York, 365 Fifth Avenue, New York, NY 10016, USA}
\email{arothermich@gradcenter.cuny.edu}

\begin{abstract}
We investigate the effects of surface gravity, effective temperature, and cloudiness on the potassium doublet (\ion{K}{1}) at 1.17 μm in brown dwarf spectra. Using pseudo-Voigt profiles to fit the \ion{K}{1} doublet in Sonora Diamondback atmospheric model spectra, we find that gravity and cloudiness affect the spectra differently in mid to late-L dwarfs. The full-width at half-maximum (FWHM) is strongly correlated with surface gravity, while the maximum depth strongly correlates with cloudiness. This method allows us to separate the effects of clouds and surface gravity on the \ion{K}{1} 1.17 μm doublet. We also find that the FWHM and maximum depth of the doublet can help to estimate the effective temperature and surface gravity in early to mid-L dwarfs.

\end{abstract}



\section{Introduction} \label{sec: intro}

Alkali lines like potassium doublets' (\ion{K}{1}) absorption strength correlate with gravity, making them useful age indicators in L dwarfs \citep{Cruz09, Aller13, Martin17}. However, research is ongoing on the effect of clouds on these features. \cite{Suarez23} showed that clouds greatly affect the mid-infrared spectra of L dwarfs, and research suggests the same is true for the \ion{K}{1} in the near-infrared, e.g. \cite{Martin17}. 
In this study, we investigate how clouds and gravity shape near-infrared \ion{K}{1} lines at different temperatures.

\section{Fitting the Pseudo-Voigt} \label{sec: method} 

We fit the \ion{K}{1} at $1.17$ μm with a pseudo-Voigt profile ($pV$). We did not study the \ion{K}{1} $1.25$ μm doublet because it is blended with \text{FeH} molecular absorption \citep{Cushing05}. A pseudo-Voigt profile is a linear combination of Gaussian and Lorentzian functions (Equations ~\ref{eq:pV}, ~\ref{eq:G}, and ~\ref{eq:L} respectively), where the Gaussian fits the core of the line and the Lorentzian captures the broadening edges.

\begin{equation} \label{eq:G}
  G(x; σ, μ) = \frac{1}{σ \sqrt{2 π} }e^{-\frac{1}{2} (\frac{x-μ}{σ})^2}
\end{equation}
\begin{equation} \label{eq:L}
 L(x; γ, μ) = \frac{γ}{2 π} \frac{1}{(x-μ)^2 + (γ/2)^2}
\end{equation}
\begin{equation} \label{eq:pV}
  pV(x; η, A, Γ, μ) = I(η, A, Γ)\left[ η G(x; σ(Γ), μ) + (1-η)L(x; γ(Γ), μ) \right]
\end{equation} 

$A$ is the maximum depth, $Γ$ is the full-width at half-maximum (FWHM), $μ$ is the center of the distribution, and $η$ is the mixing coefficient ($0 \le η \le 1$). These variables are manipulated by the functions $σ(Γ), γ(Γ)$, and $I(η, A, Γ)$ (Equations \ref{G sigma}, \ref{L gamma}, and \ref{pV scaling}) in the pseudo-Voigt profile. 

\begin{align} \label{G sigma}
σ(Γ) &= \frac{Γ}{\sqrt{2 \log(2)}} \\
\label{L gamma}
γ(Γ) &= 2Γ \\
\label{pV scaling}
I(η, A, Γ) &= A \left[ \frac{η}{σ(Γ) \sqrt{2 \pi}} + (1 - η)\frac{2}{\pi γ(Γ)} \right]^{-1}
\end{align}

Further details on the fitting method and the Python package used are available in \cite{DoubQuant}.

\section{Analysis} \label{sec: analysis}
Knowing precise surface gravity, $f_{sed}$, and temperature is essential for analyzing their effect on the \ion{K}{1} doublet, but it is challenging in observed spectra. Fortunately, atmospheric models have these parameters.

We use cloudy Sonora Diamondback models \citep{Morley24} with temperatures $1200-2400 \text{ K}$ (corresponding to L dwarfs), surface gravity in the range $\log g=3.5-5.5$, and cloud parameters ($f_{sed}$) from 1 (thick clouds) to 8 (thin clouds). We used solar metallicity for this analysis.

\begin{figure}[ht!]
  \centering
  \gridline{\fig{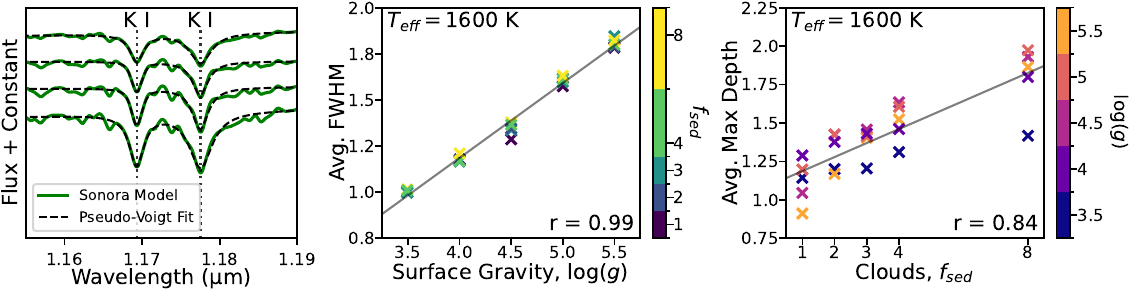}{0.9\textwidth}{}}
  \vspace{-.7cm}
  \gridline{\fig{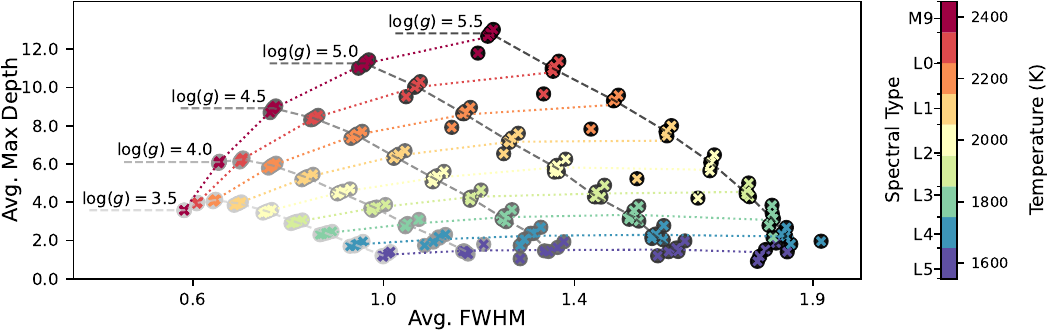}{0.9\textwidth}{}}
  \caption{
  \textbf{Top Left:} pseudo-Voigt fits (black dashed lines) to the \ion{K}{1} doublet at $1.17$ μm from Sonora Diamondback model spectra convolved to \(R\approx1000\) (green solid lines).
  \textbf{Top Middle and Right:} average FWHM vs. gravity (middle plot) and average maximum depth vs. $f_{sed}$ (right plot) for model spectra with \(T_{\text{eff}} = 1600 \text{ K}\) (L5 spectral types). The Pearson correlation coefficient is shown in the bottom right corner. 
\textbf{Bottom:} average maximum depth versus average FWHM for the \ion{K}{1} doublet. The colors of `x' indicate temperature, and the circle shades indicate gravity. Dotted lines connect points of the same temperature (by color) and the same gravity (by shade). 
  }
  \label{fig}
\end{figure}

We convolved the each of selected 325 Diamondback models to \(R=1000\) at 1.17 μm, comparable to resolutions of instruments like JWST/NIRSpec, Keck/NIRSPEC, and IRTF/SXD. This resolution allows us to fully resolve the shape of the doublet. Subsequently, we fit the pseudo-Voigt to each model using the method shown in the top left panel of Figure~\ref{fig}. The code to reproduce these results is available on GitHub\footnote{\url{https://github.com/BDNYC/quantifying_sonora_KI_doublet}}.

\section{Results} \label{sec: results}
We found the average $A$ and FWHM of both doublet absorptions are key to analyzing the effect of surface gravity, temperature, and clouds.

\subsection{Influence of Gravity and Clouds with Constant Temperature} \label{subsec: const temp}
Figure \ref{fig} illustrates the relationship between average FWHM and surface gravity (top middle panel), and average maximum depth and cloud parameter, $f_{sed}$ (top right panel) at \(1600 \text{ K}\). This temperature corresponds to mid-L types dwarfs, which exhibit a wide variety of clouds and the thickest clouds on average \citep{Suarez22}.


We observed a strong positive correlation (Pearson correlation coefficient of $r = 0.84$) between the absorptions' depth and cloudiness. This indicates that the doublet is stronger in the spectra with thin clouds (high $f_{sed}$ values). Conversely, surface gravity and maximum depth are weakly correlated or uncorrelated ($r = 0.16$).

We found the FWHM strongly correlates ($r = 0.99$) with surface gravity, where the doublet is broader for higher surface gravities. Conversely, the FWHM and cloudiness are uncorrelated ($r = 0.05$). 

\subsection{Influence of Gravity and Clouds at Various Temperatures} \label{subsec: temp constraints}
The bottom plot in Figure \ref{fig} shows the relationships between surface gravity and temperature with the shape of \ion{K}{1} doublet. Temperature follows positively curved diagonals (colored dotted lines), with steeper lines representing higher temperatures and flatter lines indicating cooler models. Surface gravity forms negative diagonals (grayscale dashed lines). The distinct trends of temperature and surface gravity make a model's position predictive of its temperature and gravity.

 The FWHM shows a strong linear correlation with gravity ($r > 0.95$) across temperatures $1400 \text{ K} \lesssim T_\text{eff} \lesssim 2400 \text{ K}$. The FWHM remains uncorrelated with cloudiness ($r \leq 0.2$), making it a reliable gravity indicator independent of cloud effects. 
 
 Similarly, the linear correlation between maximum depth and cloudiness holds ($r > 0.70$) for $1200 \text{ K} \lesssim T_\text{eff} < 1700 \text{ K}$, while maximum depth and surface gravity remain minimally correlated ($r < 0.2$). This temperature range aligns with the observed formation and sedimentation of silicate clouds ($1300 \text{ K} \lesssim T_\text{eff} \lesssim 2000 \text{ K}$) \citep{Suarez22}. These results indicate maximum depth is a good cloudiness indicator independent of gravity for models cooler than $1700 \text{ K}$. 

At higher temperatures ($T_\text{eff} > 1800 \text{ K}$), the maximum depth becomes highly correlated ($r > 0.89$) with surface gravity, and less correlation with clouds ($r < 0.30$). The correlation of the maximum depth and the FWHM with surface gravity at higher temperatures causes the increased steepness observed in the bottom plot of Figure \ref{fig}, helping to further separate the temperature and gravity clusters.

Variation in cloud abundance (\(f_{sed}\)) causes data spread at constant temperatures and gravities in the bottom plot of Figure \ref{fig}. This spread remains limited at low gravities, where points of the same temperature cluster together, indicating that clouds have minimal influence. At higher gravities, the cloudiest models align with lower temperatures and gravities, emphasizing the impact of clouds on spectra. This effect peaks in dense low L-dwarfs, where the cloudiest models are closer to models over 100 K cooler with lower gravity than their cluster; however, in real observations, the temperature constraints could be refined using other methods.

\section{Summary} \label{sec: summary}
We used pseudo-Voigt profiles to fit the \ion{K}{1} doublet at $1.17$ μm in brown dwarf spectra to quantify both the broadening and depth of the lines. We applied this method to Sonora Diamondback cloudy models. We could separate the effects of clouds and surface gravity for temperatures in the range $1400 \text{ K} \lesssim T_\text{eff} \lesssim 1700 \text{ K}$, as the example top middle and right plots of Figure \ref{fig} demonstrate.

Additional results included the non-negligible effect of clouds in hot L-dwarfs with high gravities. Consequently, clouds must be considered when unexpectedly shallow and narrow \ion{K}{1} doublets in L dwarfs are observed (Section~\ref{sec: intro}).

This method may be helpful in distinguishing cloudy brown dwarfs from low-gravity brown dwarfs with similar temperatures or spectral types. Another application is measuring the effect of spin axis inclination on the near-infrared spectra of brown dwarfs, as clouds and inclination are strongly correlated where clouds are more opaque at the equator than at the poles \citep{Suarez23}.

We also found that temperature and gravity can be estimated based on the doublet maximum depth and FWHM for $1600 \text{ K} < T_\text{eff} \lesssim 2400 \text{ K}$ as shown in the bottom plot in Figure \ref{fig}.

\software{Doublet\_Quantifier \citep{DoubQuant},
Quantifying\_Sonora\_KI\_Doublet \url{https://github.com/BDNYC/quantifying_sonora_KI_doublet}}

\bibliography{bibliography}{}
\bibliographystyle{aasjournal}

\end{document}